\documentclass[lettersize,journal]{IEEEtran}

\usepackage{amsmath,amsfonts}

\usepackage[linesnumbered,ruled,lined]{algorithm2e}
\usepackage{booktabs}
\usepackage{subcaption}
\usepackage[inkscapelatex=false]{svg}
\usepackage{multirow}
\usepackage{amsmath}
\usepackage{seqsplit}
\usepackage{csquotes}
\usepackage{colortbl}  %彩色表格需要加载的宏包
\usepackage{xcolor}
\definecolor{lb}{HTML}{E6E6E6} 
\usepackage{graphicx}
\usepackage{pifont}
\usepackage{makecell}
\usepackage{textcomp}
\usepackage{stfloats}
\usepackage{verbatim}
\usepackage{url}
\usepackage{setspace}
\usepackage{enumitem}
\usepackage{booktabs} 

\usepackage{amssymb}
\usepackage[normalem]{ulem}
\usepackage{tabularx}
\usepackage{tcolorbox}
\usepackage{cite}
\tcbuselibrary{skins}
\usepackage{threeparttable}
\usepackage{hyperref} %生成引用链接，注：该宏包可能与其他宏包冲突，故放在所有引用的宏包之后
\usepackage{cleveref} %实现图片和表格、公式的引用，cleveref包必须放在hyperref包之后
\newcommand{\circled}[1]{\ding{\numexpr171+#1\relax}}

\hypersetup{
    colorlinks = true,
    linkcolor = blue,
    urlcolor  = blue,
    citecolor = red
}

\newcommand{\salt}[0]{\textsc{ALT}\xspace}
\newcommand{\saltm}[0]{ALT4Decompile\xspace}
\newcommand{\salte}[0]{\textsc{ALT4EXE}\xspace}

\begin{document}

% ============================================================
% Title
% ============================================================

\title{\saltm: Inferring C-aligned Abstract Loop Tree for LLM-Based Binary Decompilation}

% ============================================================
% Authors
% ============================================================

\author{
Yongpan~Wang,
Puzhuo~Liu,
Xin~Xu,
Siyuan~Li,
Yaowen~Zheng,
Xiaodong~Gu,
and~Beijun~Shen%
\thanks{Yongpan Wang, Xiaodong Gu, and Beijun Shen are with
Shanghai Jiao Tong University, Shanghai, China.
E-mail: \{frankile, xiaodong.gu, bjshen\}@sjtu.edu.cn.}%
\thanks{Puzhuo Liu is with Tsinghua University, Beijing, China.
E-mail: liupz@mail.tsinghua.edu.cn.}%
\thanks{Xin Xu is with The Hong Kong University of Science and Technology,
Hong Kong SAR, China.
E-mail: xxuca@connect.ust.hk.}%
\thanks{Siyuan Li is with the School of Cyber Science and Technology,
Shandong University, Jinan, China.
E-mail: siyuan@sdu.edu.cn.}%
\thanks{Yaowen Zheng is with the Institute of Information Engineering,
Chinese Academy of Sciences, Beijing, China.
E-mail: zhengyaowen@iie.ac.cn.}%
\thanks{Corresponding author: Xiaodong Gu.
E-mail: xiaodong.gu@sjtu.edu.cn.}%
}

% ============================================================
% Paper Header
% ============================================================

\markboth{IEEE Transactions on Dependable and Secure Computing}%
{Wang \MakeLowercase{\textit{et al.}}: Inferring C-aligned Abstract Loop Tree for LLM-Based Binary Decompilation}

\maketitle

% ============================================================
% Abstract
% ============================================================

\begin{abstract}

Decompilation refers to the process of recovering high-level (C) language code from low-level (assembly) code.
Recent Large Language Model (LLM)-based methods can generate re-executable code but struggle with the enormous syntax gap between assembly and C languages.
Loop structures in C are fragmented into complex jumps within assembly languages, which poses a challenge to LLMs decompiling assembly language, which is mainly pre-trained on C.

To bridge this gap, we propose \saltm, which refactors assembly to closely resemble C language structure, thereby preventing hallucinations in LLMs caused by complex jumps.
Specifically, we construct a C-aligned \underline{\textbf{A}}bstract \underline{\textbf{L}}oop \underline{\textbf{T}}ree (\salt) based on specific jump patterns of Assembly that explicitly aggregates fragmented Assembly blocks into corresponding high-level loop structures (e.g., nested loops).
Finally, we fine-tune an LLM adapted to \salt to generate decompiled code and then improve its output by fixing specific errors and restoring symbols.
Evaluated against 12 baselines (including rule-based SAILR and LLM-based LLM4Decompile) on Decompile-Eval, MBPP, and ExeBench, \saltm achieves state-of-the-art correct re-executable results (even under four commonly used obfuscation techniques): i.e., a 70.4\% test case pass rate on Decompile-Eval, a 10.6\% improvement over prior best work.
Furthermore, analyses on real-world software and a user study further show its practicability in understanding binary functions.
The datasets and source code used in this paper are available at
\url{https://github.com/wang-yongpan/ALT4Decompile}.
\end{abstract}

% ============================================================
% Keywords
% ============================================================


% ============================================================
% Main Text
% ============================================================

\vspace{-5pt}
\section{Introduction}\label{sec:intro}
Decompilation aims to recover C code equivalent to a given binary executable, playing a crucial role in various reverse engineering tasks such as vulnerability discovery~\cite{vul1, vul2, vul3, vul4}, malware analysis~\cite{yakdan2016helping, mal2, mal3, mal4}, and understanding or reusing closed-source software~\cite{li2025lares, close2, close3, close4}. 
However, this task is inherently challenging due to irreversible information loss during the compilation from C to assembly languages~\cite{yuan2019b2sfinder, tang2020libdx, jiang2024binaryai}, such as non-intuitive code transformations introduced by compiler optimizations (e.g., the loss of C structures)~\cite{hersim, wang2022jtrans, ding2019asm2vec} and the loss of readability-oriented features (e.g., variable names and function signatures)~\cite{xie2024resym, hu2024degpt}.

Directly generating recompilable and re-executable code can integrate potential dynamic analysis and testing methods~\cite{wong2025decllm, burk2022decomperson,bohme2017directed}, thereby reducing the workload of manual reverse.
Traditional decompilers, such as IDA Pro~\cite{IDApro} and Ghidra~\cite{ghidra}, primarily focus on translating Assembly code into pseudo-code to assist reverse engineers in understanding program semantics. 
However, although pseudo-code offers better readability than Assembly code, its \textit{non-executable nature limits the application of decompilation results in downstream analysis}, such as code repackaging~\cite{reiter2024automatically}, platform migration~\cite{burk2022decomperson, verbeek2020sound}, and vulnerability debugging~\cite{burk2022decomperson, bohme2017directed}.

Existing approaches still show limitations in terms of automation, reliability, and accuracy in generating recompilable and re-executable code.
Specifically, existing approaches could be classified into three categories: 
\textbf{Rule-based methods} such as Ghidra~\cite{ghidra} and SAILR~\cite{sailr} generally employ static and dynamic analysis to recover readable pseudo-code, yet their effectiveness heavily relies on domain-specific expertise and manual verification. 
\textbf{Refinement-based methods} use large language models (LLMs) to refine the pseudo-code generated by rule-based methods to reduce manual work and improve readability or correctness. 
LLM4Decompile-Ref~\cite{tan2024llm4decompile}, DecLLM~\cite{wong2025decllm}, and ReySm~\cite{xie2024resym} focus on refining the output of existing decompilers (e.g., Ghidra).
However, such methods are susceptible to decompiler-induced binary execution logic distortion~\cite{dramko2024taxonomy, zhou2025fidelitygpt} and require considerable cost to adapt to new architectures~\cite{IDApro, ghidra}.
\textbf{End-to-end methods} address the above limitations by processing Assembly code directly.
Nova~\cite{jiang2023nova} fine-tuned DeepSeekCoder~\cite{guo2024deepseek} using hierarchical attention.
% on a large-scale Assembly-source dataset.
LLM4Decompile-End~\cite{tan2024llm4decompile} trains an end-to-end decompilation model on a large-scale Assembly-C pairs dataset from ExeBench~\cite{armengol2022exebench}.
SccDec~\cite{sccdec} enhances it by introducing a self-constructed context from compilation.
These methods show great potential, but still face recompilable and re-executable problems.

\begin{figure*}
    \centering
    \includegraphics[width=\linewidth]{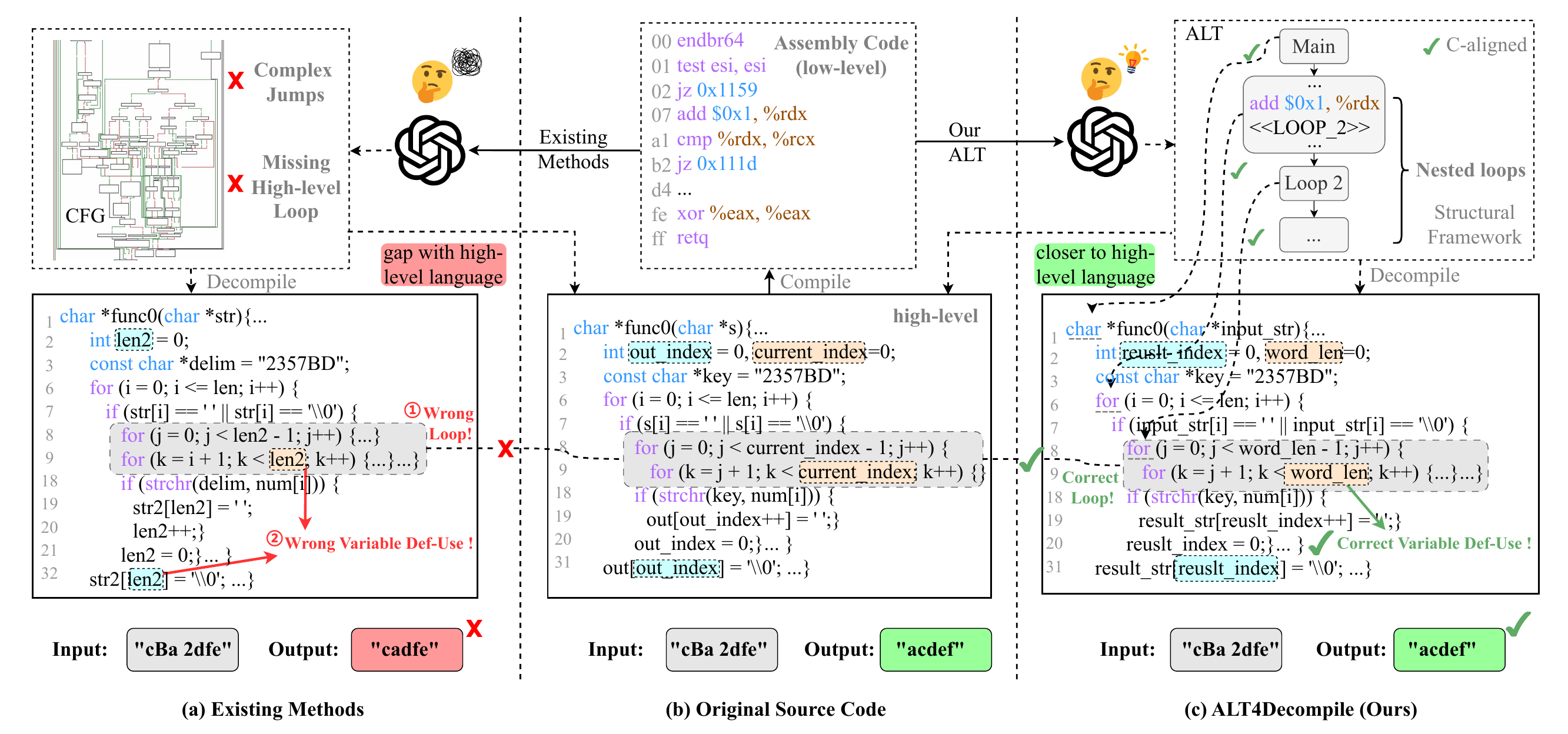}
    %\vspace{-2em}
    \caption{A Motivation Example from Decompile-Eval. IN means instruction normalization.}
    \label{fig:me1}
   % \vspace{-1.5em}            
\end{figure*}

Specifically, existing methods simply treat Assembly as a high-level language, relying entirely on LLMs' inherent capabilities while ignoring the critical syntax gap between high- and low-level languages.
During compilation, loop structures in C are fragmented into complex instruction jumps.
This fragmentation challenges LLMs that are primarily pre-trained on C languages.
Although control flow graphs (CFGs) provide detailed jump flows, the loop structure of C is invariably fragmented across multiple basic blocks of CFG.
LLMs struggle to grasp these scattered relationships to correctly reconstruct structures like loops.
Consequently, complex control flows frequently induce structure and variable errors (shown in~\circled{1} and~\circled{2} of Figure~\ref{fig:me1}(a)), leading to execution behavior deviations.

In this paper, we present \saltm, a novel LLM-based binary decompilation technique that refactors Assembly into a C-aligned \underline{\textbf{A}}bstract \underline{\textbf{L}}oop \underline{\textbf{T}}ree (\textbf{\salt}, Section~\ref{sec:salt}).
For complex jumps of assembly, \salt initially constructs a \textbf{structural framework from the C perspective} rather than relying solely on LLMs' inherent capabilities.
Specifically, the loop structure of C languages and some specific jump patterns of assembly languages share common logic features (e.g., repeated executions indicating loops).
By analyzing specific jump patterns, we aggregate instructions from the same C loop into unified, hierarchically organized nodes.
These nodes explicitly define loop boundaries aligned with C structure syntax, successfully mitigating LLMs' struggles with complex control flows (e.g., on the Decompile-Eval benchmark, our improvement increases from 22.8\% on samples without nested loops to 26.4\% on samples with nested loops).
This aggregation and layering mechanism enables LLMs to initially identify the program structure among complex jumps, thereby preventing disorientation and reducing hallucinations.
To generate decompiled code, we fine-tune an LLM adapted to \salt and improve its output by fixing specific errors and restoring symbolic information.

We evaluate \saltm on three public datasets (Decompile-Eval, MBPP, Exebench) with input/output (I/O) samples and compare \saltm against three categories of baselines, i.e., general-purpose LLMs, rule-based decompilers, and LLM-based decompilers (including refinement-based and end-to-end methods).
The results show that \textit{\saltm achieves new state-of-the-art (SOTA) performance}. 
For example, on the Decompile-Eval dataset, \saltm achieves a recompilation rate of 96.8\% (4\% improvement over SOTA), a re-execution rate of 58.7\% (8.9\% improvement over SOTA), and a test case pass rate of 70.4\% (10.6\% improvement over SOTA).
Furthermore, \saltm consistently outperforms baselines across four common code obfuscation techniques, real-world software (e.g., WeChatWin.dll), and model types. 
A user survey further confirms that our decompilation output is the preferred choice over competing methods.

In summary, our contributions are as follows:

\begin{itemize}[leftmargin=10pt]
% \setlength{\parindent}{0pt}
%\item To the best of our knowledge, \saltm is the first approach that mitigates hallucinations in LLM-based binary decompilation caused by complex control jumps, leveraging stable abstract logic features shared between binaries and source code.
%And we release the model weights and code on the website~\cite{SALT4Decompile} to support future research.
\item To the best of our knowledge, \saltm is the first approach to bridge the syntactic gap (e.g., complex instruction-level jumps) between low- and high-level languages by leveraging their shared logic features, thereby improving LLM-based decompilation performance.
And we release the model weights and code on the website~\cite{SALT4Decompile} to support future research.
\item We propose the first practical algorithm to construct \salt from Assembly code, demonstrating its feasibility and superiority in assisting LLM-based decompilation by establishing a structural framework from the high-level language perspective.
%We first explore the superiority and feasibility of extracting \salt from Assembly code and propose a practical algorithm for this extraction, which can assist in LLM-based binary decompilation through constructing a logic framework from the perspective of high-level language.
\item We conduct extensive experiments (e.g., code obfuscation and real-world software) to assess the effectiveness of our method. 
The experimental results show that \saltm achieves new SOTA performance in binary decompilation, attaining a 70.4\% test case pass rate on the Decompile-Eval dataset, representing a 10.6\% improvement over the previous best approach. 
\end{itemize}

%Our source code and experimental data for reproducibility are publicly available on the website\footnotemark[\value{footnote}]}. 

\section{Motivation}\label{sec:mt}

\smallskip\noindent\textbf{Problem.}
Compared to traditional rule-based decompilers, LLM-based decompilers promise automatically executable outputs, supporting downstream reverse engineering. However, existing LLM solutions~\cite{tan2024llm4decompile, sccdec, liupeipei} conservatively treat Assembly as a new high-level language, ignoring its unique low-level syntax.
Unlike C, Assembly lacks explicit high-level loop structures, relying instead on complex jumps (e.g., JMP $<$address$>$) of instructions.
LLMs, which are pretrained primarily on high-level language, struggle to reconstruct relationships between scattered instructions and translate them into corresponding loop structures of C.

Figure~\ref{fig:me1} illustrates it using some simplified \seqsplit{Decompile-Eval}~\cite{tan2024llm4decompile} examples.
The source code (Figure~\ref{fig:me1}b) features a three-level nesting loop removing characters specified by the constant $key$ and sorts the remainder (e.g., \enquote{cBa2dfe} becomes \enquote{acdef}).
We use LLM4Decompile~\cite{tan2024llm4decompile} (Figure~\ref{fig:me1}a) for comparison, which is the first and most widely adopted approach. 
LLM4Decompile incorrectly generates a two-level loop and merges the definitions and usages of two distinct variables ($current\_index$, $out\_index$) into one ($len2$), causing failure.
In our manual analysis, excluding failures caused by missing corresponding content such as data, over 80\% of the failures arise from complex jumps induced by loops.
% Second, due to isolated data segments, LLM4Decompile misses the $key$ value reference, ultimately failing to filter \enquote{B} and \enquote{2}.

\smallskip\noindent\textbf{Insight.}
To tackle these problems, we must construct valuable information about C structures from Assembly to guide the LLM.
Simply relying on LLMs' inherent capability to infer C code execution logic from raw Assembly or CFGs is challenging, because LLMs are pretrained primarily on high-level language.
However, converting Assembly to C-like pseudo-code scales poorly and risks execution logic distortion~\cite{dramko2024taxonomy, zhou2025fidelitygpt}.
Therefore, we aim to refactor Assembly into a structure resembling C while retaining original execution logic in the corresponding structure. 
Execution logic without complex jumps is easier for LLMs to process.
Inspired by prior works~\cite{IDApro,angr, ghidra}, we observe that some specific jump patterns of Assembly code mirror C loop structures (i.e., repeat the execution of the same code segment).
% \smallskip\noindent\textbf{\salt.}
Based on this insight, we propose \salt (Section~\ref{sec:salt}), inferred directly from Assembly. \salt is a rooted tree $T = (N, E, r, \mathcal{I})$, where $N$ represents loop structure nodes; $E \subseteq N \times N$ defines parent-child relationships (e.g., nested sub-loops); $r \in N$ is the function body root; and $\mathcal{I}$ maps each node to an ordered sequence of corresponding Assembly instructions.

\smallskip\noindent\textbf{Why \salt is effective?}
It equips LLMs with a high-level language structural framework inferred via simple shared logic features.
Unlike CFG-based methods~\cite{liupeipei} forcing LLMs to implicitly infer C structures from scattered blocks, \salt explicitly aggregates instructions belonging to the same C loop into unified nodes.
This establishes clear boundaries mirroring C structure syntax.
As Figure~\ref{fig:me1}c shows, every instruction that belongs to the same C loop structure is aggregated into a single node, thereby producing a well-structured, high-level logic representation with clear boundaries.
This simplified structure effectively guides the LLM to achieve accurate loop structure recovery.
Even when a small fraction of loop nodes with unidentifiable structural boundaries are aggregated, the internal jumps within these nodes remain simpler than the global control flow of the original code, which facilitates inference for LLMs.
Notably, our empirical analysis shows that ALT's relative improvement over the baseline \textbf{increases from 22.8\% on samples without nested loops to 26.4\% on samples with nested loops}. 
In addition, Table~\ref{table:ablation_salt} and Section~\ref{ab_1} show that the performance of our \salt (57.2\% RE) is greater than CFG (51.8\% RE) for decompilation. 
% In addition, \salt leverages architecture-independent logic features~\cite{binenhance, jiang2024binaryai, tang2020libdx}.
Unlike intermediate representations (IRs), \salt is exclusively designed to generate re-executable decompiled code.

\section{Design of \saltm}

\begin{figure*}
    \centering
    %\vspace{-2em}
    \includegraphics[width=\textwidth]{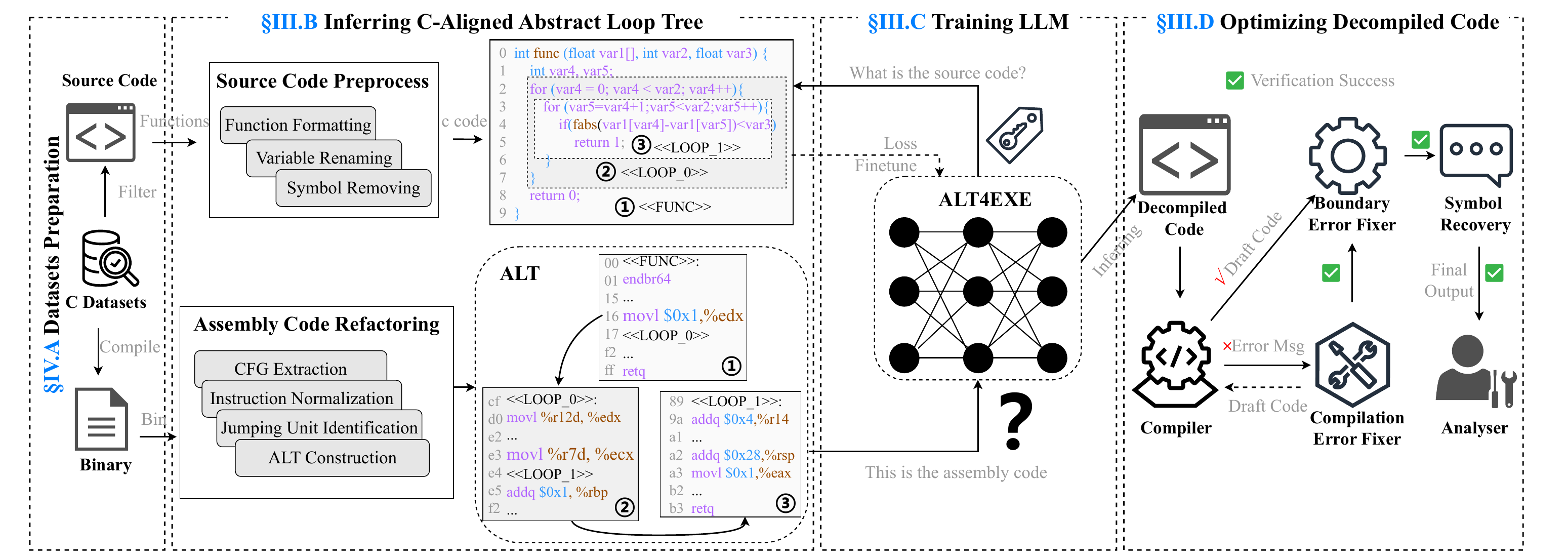}
    %\vspace{-2em}
    %\vspace{12.5em}
    \caption{The Workflow of \saltm.}
    \label{fig:workflow}
    %\vspace{-1.5em} 
\end{figure*}

\subsection{Overview}
The core idea of our method is to construct the abstract high-level structures (i.e., nested loops) from low-level binary operations, which enables LLMs to better understand the relationships among scattered basic blocks.
Inspired by prior works~\cite{ghidra, IDApro, angr}, we derive these structures by analyzing specific jump patterns, such as the repeated execution of a given code segment.
Figure~\ref{fig:workflow} illustrates the overall workflow of \saltm, comprising three key stages.
First, we construct C-aligned abstract loop trees (\textbf{\salt}) from input binaries (Section~\ref{sec:salt}).
Next, we fine-tune a language model (\textbf{\salte}) using the \salt to generate decompiled code (Section~\ref{sec:training}).
Finally, we refine the output by correcting predefined errors while recovering symbolic information (Section~\ref{sec:op}).

\subsection{Constructing C-Aligned Abstract Loop Tree}\label{sec:salt}

Binary code often contains complex jump instructions that disrupt statement coherence, impeding language models from accurately understanding high-level code logic. 
To overcome this, we design \salt, a novel abstract binary representation that leverages shared logic features between low- (Assembly) and high-level (C) languages. 
For instance, loop structures consistently manifest as cycles with back-edges in CFGs across both levels.
Thus, \salt reconstructs the initial high-level loop framework directly from Assembly code.

Specifically, \salt is a tree structure encapsulating jump structures like loops. 
The root node represents the entire function body.
At each jump target, a special \textit{marker instruction} is inserted to indicate a jumping unit (Figure~\ref{fig:salt}).
The root's child nodes represent first-level jumping units (e.g., Loop\_0 and Loop\_2 in Figure~\ref{fig:salt}) within the corresponding high-level language, named after their corresponding markers. 
This parent-child relationship recursively descends to leaf nodes, which represent atomic, jump-free instruction blocks.
Each node strictly orders its Assembly instructions by address.
The formal definition of \salt is detailed in Section~\ref{sec:mt}.
The recovery process of \salt consists of CFG extraction, instruction normalization, jumping unit identification, and tree construction.

\subsubsection{CFG Extraction}
%SALT can be inferred from the control flow graph (CFG) of a binary function.
We start by extracting the CFG from the binary code, which is widely used in binary analysis~\cite{isrd, gemini}. 
Nodes of a CFG represent basic blocks composed of multiple Assembly instructions, while edges represent possible control flow transitions between these blocks.
Although the CFG partially captures program logic~\cite{liupeipei}, LLMs still fail to accurately model complex control flow, leading to unstable decompilation results (detailed in Section~\ref{ab_1}).

\subsubsection{Instruction Normalization (IN)}\label{sec:in}
Differences in data storage mechanisms between low-level and high-level languages necessitate the extraction of isolated data segments based on their absolute addresses.
Following this extraction, we normalize the instructions within the CFG basic blocks using two distinct approaches.
First, for jump instructions (e.g., $je$), we convert absolute addresses to relative offsets from the function's entry point.
Second, for instructions referencing data segments (e.g., .data, .rodata), we extract the corresponding content (i.e., convert them to strings) and append it to the instructions as comma-separated Assembly comments.

\subsubsection{Detect Jumping Units}
We identify jump units from the normalized CFGs and prioritize loops due to their structural significance in defining execution paths and affecting decompilation correctness, as Section~\ref{sec:mt} demonstrates.
We defer the incorporation of other components, such as conditional structures, to future investigations.
Loops consistently manifest as back edges in CFGs, directing control flow to repeat code fragments and forming strongly-connected subgraphs.
Leveraging this structure~\cite{angr, ghidra, IDApro}, we identify loop blocks in the binary's CFG corresponding to high-level language loops and extract their nested hierarchies.

\begin{algorithm}
    \small
    \setstretch{1.0}
    \SetAlgoLined
    \caption{Jumping Unit Identification}
    \KwIn{Control Flow Graph $G$}
    \KwOut{Nested Loop Structures $ls$}
    \label{algo:a1}
    \BlankLine
    $ls \leftarrow \emptyset$\;
    \SetKwProg{Def}{Function}{:}{end}
    \SetKwFunction{}{}
    \SetKwFunction{}{DetectLoop}
    
    \Def{DetectLoop($G$)}{
    \SetAlgoVlined
        $sub\_gs \leftarrow \text{GetConnectedGraphs}(G)$\;
        \While{$sub\_gs \neq \emptyset$}{
            $g \leftarrow sub\_gs.\text{pop}()$\;
            $loop \leftarrow \text{CreateNewLoop}(g, ls)$\;
            \For{$sub\_g \in sub\_gs$}{
                \If{$sub\_g\in g$}{
                    $nl \leftarrow \text{CreateNewLoop}(sub\_g, ls)$\;
                    $loop.children$\text{.append($nl$)}\;
                    $ls = ls\cup\{nl\}$
                }
            }
                $ls = ls\cup \{loop\}$\;
        }
        \Return $ls$\;
    }
\end{algorithm}

\begin{figure*}
    \centering
    %\vspace{-2em}
    \includegraphics[width=\textwidth]{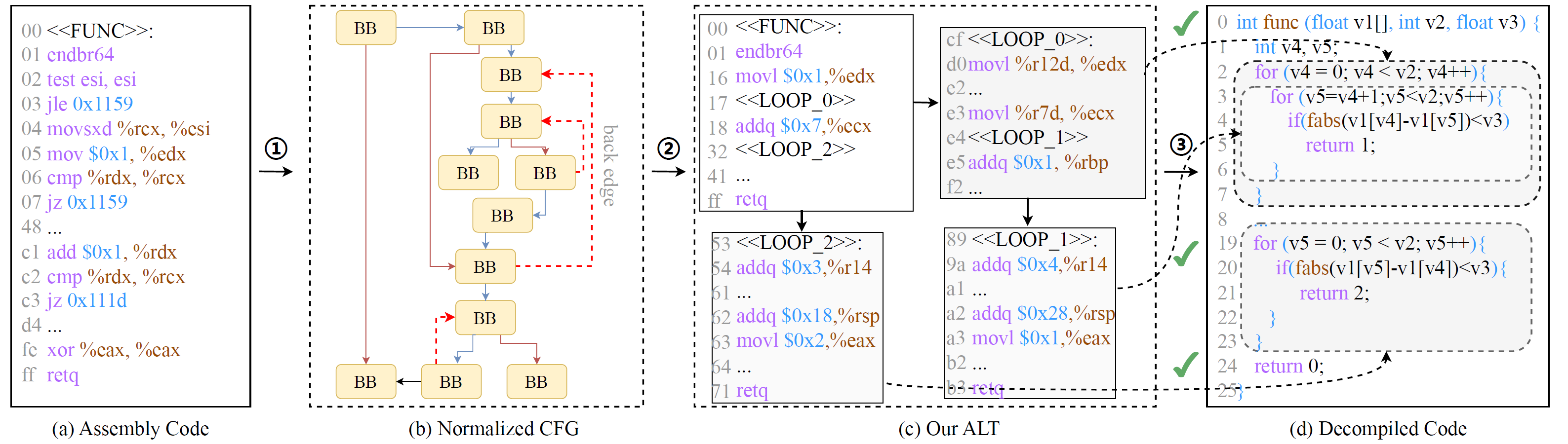}
   % \vspace{-2em}
    \caption{The Process for Constructing \salt. Grey boxes represent the identified loop structures.}
    \label{fig:salt}
    %\vspace{-1.5em} 
\end{figure*}

% Algorithm~\ref{algo:a1} extends the classic CFG-based loop detection method~\cite{angr} with improvements like multi-entry handling (e.g., randomly select entry nodes) for \salt construction.
Inspired by previous methods~\cite{angr, ghidra, IDApro}, Algorithm~\ref{algo:a1} outlines the process of detecting loop structures from CFGs, which is crucial for constructing \salt.
The principle is to detect strongly connected subgraphs and analyze their inclusion relationships to determine nesting.
Notably, we add some measures to handle situations that fail to detect, e.g., selecting a minimum in-degree entry node for multi-entry subgraphs.
First, we extract connected subgraphs via depth-first search (DFS)~\cite{dfs} (Line 3), where each subgraph corresponds to a c-aligned loop structure.
We then traverse these subgraphs to build the nested loop structure $ls$ (Lines 4-13). 
Specifically, we check if a subgraph fully contains another to identify lower-level loops (Line 8), while CreateNewLoop manages loop creation (Lines 6, 9).
Finally, the algorithm outputs the unified nested loop hierarchy, matching the gray boxes in Figure~\ref{fig:salt}(c) and (d). 

\begin{algorithm}
    \small
    \setstretch{1.0}
        \SetAlgoLined %显示end
	\caption{\salt Construction}%算法名字
	\KwIn{Control Flow Graph $G$, All loops $ls$, entry node $en$, parent block $pb$}%输入参数
	\KwOut{C-aligned Abstract Loop Tree $alt$}%输出
    \label{algo:a2}
	\BlankLine
         $alt \leftarrow \text{LoopBlock(func\_name)}$, $index \leftarrow 0$, $pb \leftarrow salt$\;
        \SetKwProg{Def}{Function}{:}{end}
        \SetKwFunction{}{}
        \Def{Construct\salt($G$, $ls$, $en$, $pb$)}{
        \SetAlgoVlined
            $loop \leftarrow$ \text{in\_loop}($en.addr$, $ls$)\;
            \If{$loop$ and not $loop$.processed}{
                $loop\text{.processed} \leftarrow$ \text{True}\;
                $block\_name \leftarrow$ $<<LOOP\_index>>$\;
                $index \leftarrow index + 1$ \;
                $lb \leftarrow$ \text{LoopBlock}($block\_name$)\;
                $pb$\text{.add\_children($lb$)}\;
                $lb$\text{.add\_ins($G$, $en$)} \;
                $pb$\text{.add\_ins($block\_name$, NULL)} \;
                \For{$sl \in loop$\text{.subloops}}{
                    $alt \leftarrow$\text{Construct\salt($G$, $ls$, $sl.node$, $lb$)}
                }
                $out\_nodes \leftarrow$ \text{GetOutNodes($G$, $en$, $loop$)}\;
                \For{$on \in out\_nodes$}{
                    $alt \leftarrow$\text{Construct\salt($G$, $ls$, $on$, $lb$)}\;
                }
                
            }
            \Else{
                $last\_ins \leftarrow$ \text{GetLastIns($en$)}\;
                $succ\_ns \leftarrow$ \text{GetSuccessors}($en$.addr)\;
                $pb$\text{.add\_ins($G$, $en$)} \;
                \For{$sn \in succ\_ns$}{
                    \If{\text{IsCall($last\_ins$)}}{
                        $alt \leftarrow$\text{MergeNodes($G$, $ls$, $sn$, $pb$)}
                    }
                    \Else{
                       $alt \leftarrow$\text{Construct\salt($G$, $ls$, $sn$, $pb$)} 
                    }
                }
            }
            \textbf{return} $alt$\;
        }
        \BlankLine
        \SetKwProg{Functionl}{Function}{:}{end}
        \Functionl{MergeNodes($G$, $ls$, $en$, $pb$)}{
        \SetAlgoVlined
            $pb$\text{.add\_ins($G$, $en$)} \;
            $succ\_ns \leftarrow$ \text{GetSuccessors($en$.addr)}\;
            \For{$succ\_n \in succ\_ns$}{
                $alt \leftarrow$\text{Construct\salt($G$, $ls$, $succ\_n$, $pb$)} 
            }
            \textbf{return} $alt$\;
            
        }
\end{algorithm}

\subsubsection{Tree Construction}\label{sec:sr}
Next, we construct an abstract tree based on the identified nested loops, as described in Algorithm~\ref{algo:a2}. 
The process begins by initializing the loop tree with a root block labeled the function's name or address.
We then recursively traverse the CFG from the entry node to identify nodes belonging to the same loop block. 
For nodes within a loop structure (Line 4-16), we create a new logic block named \seqsplit{$<<LOOP\_index>>$} as the current block, where $index$ increases with each new block.
All nodes belonging to the loop are merged into this block, which is then added as a child node to the upper-level logic block (initially from the root). 
The block's name is inserted as a special instruction in the parent logic block to mark the loop's position.
Sub-loops (Lines 12-13) and exit nodes (identified using the function GetOutNodes, Line 14) are processed recursively until all CFG nodes are processed.
To address compiler optimizations that merge identical basic blocks, we deduplicate Assembly instructions based on their addresses to avoid redundancy.
For each node outside a loop structure (Lines 17-25), we collect its Assembly instructions along with its list of child nodes.
If a node ends with a call instruction (function IsCall, Line 22), it is merged with its child nodes (function MergeNodes, Line 28), and the instructions are added to the parent logic block in address order. This process is applied recursively to all descendant nodes.
Finally, the algorithm outputs a C-aligned abstract logic tree, which is essential for training our decompilation model.

Figure~\ref{fig:salt} shows the \salt construction processing using a concrete example.
%The source code in Figure~\ref{fig:salt} (a) contains a two-level nested loop and a one-level loop, compiled with optimization level \texttt{O2}. 
For the given binary, we first disassemble it (Figure~\ref{fig:salt}(a)) and extract its CFG.
We then apply two normalization methods to the instructions within each basic block of the CFG.
% After instructions normalization, 
Following normalization, we identify back edges (highlighted as red dashed lines in Figure~\ref{fig:salt}(b)) to infer the C-aligned loop structures.
Finally, we construct the \salt (Figure~\ref{fig:salt} (c)), composed of loop blocks.
The root block represents the entire function body, while the remaining blocks correspond to the identified loops.
The hierarchical structure of the \salt reflects the nesting relationships among these loops (Figure~\ref{fig:salt} (d)).

\subsection{Training Decompilation Model}\label{sec:training}
The constructed \salt provides a logically coherent representation that is similar to the source code.
Our subsequent objective is to train an LLM to generate the decompiled code using our \salt.
To serialize \salt, node instructions are organized by address, with nodes themselves arranged in parent-child order.
Details of the training dataset are shown in Section~\ref{sec:rn}.
We adopt a sequence-to-sequence framework to train our decompilation LLM, namely \salte.
Specifically, we initialize \salte using checkpoints from LLM4Decompile~\cite{tan2024llm4decompile} and fine-tune it using the constructed pairs of (\salt, preprocessed Source Code).
Following established practices in previous work~\cite{tan2024llm4decompile, sccdec, jiang2023nova}, we employ the same training template in the training phase. 
In addition, we discard all function pairs that exceed the maximum input length of the model.

\subsection{Optimizing Decompiled Code}\label{sec:op}

\begin{table*}[t]
    \centering
    \caption{Prompt Templates for Components of \saltm}
    \label{tab:prompt}
    \small

    % Make X columns vertically centered
    \renewcommand{\tabularxcolumn}[1]{m{#1}}

    \begin{tabularx}{\textwidth}{
        >{\centering\arraybackslash}m{2.8cm}
        >{\raggedright\arraybackslash}X
    }
        \toprule
        \textbf{Component} &
        \centering\arraybackslash\textbf{Prompt Template} \\
        \midrule

        \saltm &
        This is the assembly code:\verb|\n| \{\salt\}\verb|\n| What is the source code? \\
        \midrule

        Compilation Error Fixer &
        Please fix the following code based on the error messages provided by the GCC compiler to ensure successful compilation. The fix should minimize changes to the original code while ensuring the correctness of its logic. \\
        \midrule

        Boundary Error Fixer &
        Analyze the following code to identify any possible boundary condition errors (The judgment statement of the loop is wrong, such as n is changed to n-1, the index overflow of the array is not reinitialized, such as i++ is not initialized to the original 0 in time, the variable i of the loop is initialized incorrectly, and so on), and ensure that the execution logic of the code is correct. If found, modify only the necessary parts and output the modified code. Do not modify any unrelated sections. No explanation. \\
        \midrule

        Variable Name Recovery &
        Help me rename the variables for the snippet in the following C code. The renaming principle is: high readability, simple and easy to understand, and frequently used. No explanation. \\
        \midrule

        Comments Generation &
        Help me add code comments for the code snippet in the following C code. The principle of adding comments is: high readability, simple and easy to understand, a simple code line does not add comments. No explanation. \\
        \bottomrule
    \end{tabularx}
\end{table*}

Finally, we optimize the decompiled code generated by the trained \salte.
The refinement process is designed primarily to ensure workflow completeness rather than to serve as a main contribution.
As demonstrated in Table~\ref{table:rq1}, \salt outperforms other baselines even when this process is omitted.
In contrast to existing refinement methods~\cite{decgpt, wong2025decllm}, we incorporate a verification mechanism after each step to prevent excessive modifications.
Specifically, we utilize an LLM to evaluate the differences between the input and output, determining whether the changes satisfy the modification requirements of that step.
We discard the changes that fail to meet these criteria.

This involves refining its execution logic using compiler feedback and correcting predefined boundary errors.
Additionally, we recover the variable names that are renamed as $var\_i$ during training from symbol recovery.
These optimizations are performed using general-purpose LLMs, guided by the prompt templates outlined on the Table~\ref{tab:prompt}.

\subsubsection{Compilation Error Fixer (CEF)}\label{sec:cef}

We first compile the decompiled code using the \textit{gcc} compiler.
If compilation fails, the error messages along with the decompiled code are fed into the CEF for correction.
The resulting code is then recompiled. 
If compilation errors persist, the fix process is repeated for up to three iterations.
Once the compilation succeeds or the maximum number of attempts is reached, the code is directly forwarded to the next fixer.

\subsubsection{Boundary Error Fixer (BEF)}\label{sec:bef}

Due to the division of logic blocks, a small number of execution errors arise from incorrect loop boundary conditions. 
To address this, we analyze common error patterns in \salte's output and design predefined prompts to guide the BEF in correcting them. 
Typical issues include misconfigured loop termination condition (e.g., using \textit{n+1} instead of \textit{n}), missing initialization of loop index variables (e.g., failing to set the index to $0$), and so on.
These types of errors can often be identified and corrected without requiring access to the source code.
%Therefore, we employ BEF to perform a straightforward correction of such errors.

\subsubsection{Symbol Recovery}

Following previous work~\cite{hu2024degpt}, we restore symbolic information to enhance the readability of the decompiled code. 
Specifically, we focus on recovering two widely used types of symbolic information: variable names and comments. This restoration is achieved by leveraging off-the-shelf LLMs, which infer meaningful variable names and generate relevant comments based on the context and logic of the decompiled code. 

\section{Evaluation}
We evaluate \saltm through automatic and human studies, and examine each component's impact on overall performance.
Specifically, we address the following research questions:
%We evaluate the performance of \saltm through both automated benchmarks and human studies, and further investigate the contribution of each component to the overall effectiveness. Specifically, we aim to answer the following research questions:

\begin{itemize}[leftmargin=10pt]
    \item \textbf{RQ1:} How effective is \saltm at decompilation compared to existing SOTA baselines?
    
    \item \textbf{RQ2:} Is \saltm still effective when handling binaries obfuscated using different techniques?  

    \item \textbf{RQ3:} What is the contribution of each component to the overall performance (e.g., \salt)?
    
    % \item \textbf{RQ3:} How much does \saltm assist with reverse engineering real-world binary software?
    \item \textbf{RQ4:} To what extent does \saltm assist with reverse real-world binary software (e.g., WeChatWin.dll)?
    
    \item \textbf{RQ5:} How much does it cost (i.e., time and money) to use \saltm for binary decompilation?

\end{itemize}
In addition, we conduct a user study to assess its practical utility for reverse engineering.

\begin{table*}[t]
    \centering
    \caption{Effectiveness of \saltm against Baselines on the Three Decompilation Datasets}
    \label{table:rq1}

    \small
    \setlength{\tabcolsep}{4pt}
    \renewcommand{\arraystretch}{1.08}

    \begin{tabular*}{\textwidth}{
        @{\extracolsep{\fill}}
        l
        ccc
        ccc
        ccc
        @{}
    }
        \toprule

        \multirow{2}{*}{\textbf{Model}} &
        \multicolumn{3}{c}{\textbf{Decompile-Eval}} &
        \multicolumn{3}{c}{\textbf{MBPP}} &
        \multicolumn{3}{c}{\textbf{ExeBench}} \\

        \cmidrule(lr){2-4}
        \cmidrule(lr){5-7}
        \cmidrule(lr){8-10}

        &
        \textbf{RE} & \textbf{RC} & \textbf{TCP} &
        \textbf{RE} & \textbf{RC} & \textbf{TCP} &
        \textbf{RE} & \textbf{RC} & \textbf{TCP} \\

        \midrule

        \rowcolor{lb}
        \multicolumn{10}{c}{\textbf{General-Purpose LLMs}} \\

        DeepSeek-V3
        & 39.5 & 86.9 & 51.6
        & 47.0 & 77.0 & 50.5
        & 16.7 & 67.1 & 19.8 \\

        GPT-4o
        & 19.2 & 81.1 & 34.1
        & 23.0 & 74.5 & 26.6
        & 15.9 & 65.2 & 18.8 \\

        Claude-3.5-Sonnet
        & 50.5 & 96.0 & 63.8
        & 48.0 & 80.5 & 53.2
        & 19.2 & 75.1 & 20.5 \\

        o1-mini
        & 18.3 & 78.4 & 33.7
        & 18.5 & 63.5 & 23.8
        & 14.4 & 64.1 & 17.2 \\

        \rowcolor{lb}
        \multicolumn{10}{c}{\textbf{Rule-Based Decompilation Tools}} \\

        Hex-Rays
        & 27.0 & 30.0 & 28.3
        & 25.7 & 28.3 & 26.4
        & 3.7 & 25.4 & 3.8 \\

        Ghidra
        & 26.7 & 48.2 & 26.9
        & 24.6 & 29.1 & 25.8
        & 3.2 & 24.3 & 4.1 \\

        RetDec
        & 32.9 & 79.1 & 40.4
        & 35.6 & 72.0 & 39.1
        & 19.9 & 51.6 & 21.1 \\

        SAILR
        & 7.2 & 30.6 & 10.1
        & 16.6 & 34.6 & 18.6
        & 8.1 & 37.4 & 12.5 \\

        \rowcolor{lb}
        \multicolumn{10}{c}{\textbf{Refinement-Based Method}} \\

        LLM4Decompile-Ref
        & 52.3 & 94.7 & 64.2
        & 39.4 & 64.9 & 41.8
        & 11.9 & 51.5 & 12.8 \\

        \rowcolor{lb}
        \multicolumn{10}{c}{\textbf{End-to-End Methods}} \\

        Nova-6.7B
        & 34.1 & 85.7 & 45.4
        & 20.0 & 65.0 & 23.2
        & 10.1 & 60.2 & 11.4 \\

        LLM4Decompile-End-6.7B
        & 46.7 & 88.0 & 57.9
        & 46.5 & 78.3 & 50.6
        & 21.5 & 78.3 & \underline{24.4} \\

        SccDec-6.7B
        & \underline{49.8} & \underline{92.8} & \underline{59.8}
        & \underline{47.6} & \underline{80.1} & \underline{52.1}
        & \underline{23.1} & \underline{84.7} & 23.9 \\

        \textbf{ALT4Decompile-6.7B}        & \textbf{58.7} & \textbf{96.8} & \textbf{70.4}
        & \textbf{52.4} & \textbf{81.1} & \textbf{56.4}
        & \textbf{26.4} & \textbf{87.1} & \textbf{27.8} \\

        ALT4Decompile-ALT$^*$
        & 57.2 & 91.3 & 67.1
        & 51.1 & 80.6 & 55.9
        & 26.1 & 87.0 & 27.4 \\

        \bottomrule
        
    \end{tabular*}
    
% \vspace{2pt}
% \begin{minipage}{0.99\textwidth}
% \footnotesize
% \textit{Notes.} * means that \saltm removes all refinement processes of Section~\ref{sec:op}.
% \end{minipage}

\begin{tablenotes}
    \footnotesize
    \item * means that \saltm removes all refinement processes of Section~\ref{sec:op}.
\end{tablenotes}
\end{table*}

\subsection{Datasets}

\subsubsection{Training Dataset}\label{sec:rn}
We construct our training dataset in three steps:
\textbf{1) Source Code Selection.} Following prior work~\cite{tan2024llm4decompile, sccdec, jiang2023nova}, we use the \textit{train\_real\_compilable} subset of Exebench~\cite{armengol2022exebench} ($\sim$700K C functions). To align with our objectives, we apply three filters: retaining functions with 5--500 non-blank lines, and ensuring successful GCC~\cite{gcc} compilation. Additionally, observing that many functions contain a high proportion of trivial assignment statements, we require a minimum structure-to-line ratio of 1:200 to filter out simple functions (while randomly keeping 20\% of structure-free functions to ensure diversity). This yields a final set of $\sim$40,000 C functions to balance training costs. 
\textbf{2) Source Code Preprocessing.} To help the \salte model focus on recovering execution logic, we strip extraneous elements. Specifically, we normalize formatting with clang-format~\cite{clang-format}, uniformly rename variables (e.g., \textit{var1}, \textit{var2}) using Tree-sitter~\cite{tree-sitter}, and remove comments and keywords like \textit{inline}.
\textbf{3) Compilation \& Disassembly.} We compile the retained functions into binaries using \texttt{gcc-8.4.0} under four optimization levels (\texttt{O0--O3}), and strip all symbols for fairness, producing $\sim$160,000 binary functions. We then disassemble them into AT\&T syntax using Capstone (aligning with LLM4Decompile's format). 
Finally, we extract \salt from the binary functions and pair them with their corresponding source code to fine-tune our LLM.

\subsubsection{Test Dataset}
Following prior work~\cite{sccdec, jiang2023nova, tan2024llm4decompile, decompilebench} like Decompile-bench, we select three recognized datasets measuring functional correctness via input/output verification:
1) \textbf{Decompile-Eval}~\cite{tan2024llm4decompile}: Contains 164 C functions (from \textit{HumanEval}), yielding 656 binaries across \texttt{O0}--\texttt{O3}, evaluated via assertion-based tests (4-5 per function).
2) \textbf{MBPP}~\cite{decompilebench}: 974 Python programming problems ported to C. Compilation under four optimization levels yields 3896 binary functions, and each with 3 assertion tests.
3) \textbf{Exebench}~\cite{armengol2022exebench}: Comprises real-world C functions sourced from GitHub. Specifically, we use its \texttt{test\_real} subset, featuring complex user-defined data structures and richer dependencies. After filtering non-compilable functions, we retain 1,575 C functions (6,300 binaries across \texttt{O0}--\texttt{O3}). In addition, we modify the test code by adding macro defintions to calculate the proportion of passed cases. 

\subsection{Experimental Setting}
\subsubsection{Baselines}
We compare \saltm with three categories of baselines.
1) \textbf{General-purpose LLMs}: DeepSeek-V3~\cite{deepseekv3}, GPT-4o~\cite{gpt-4o}, o1-mini~\cite{o1-mini}, and Claude-3.5~\cite{claude35}, evaluated via official APIs given their massive parameter sizes.
2) \textbf{Rule-based Tools}: Three widely recognized top-tier decompilers in the field~\cite{eom2024r2i} (Hex-Rays~\cite{hex2020ida}, Ghidra~\cite{ghidra}, RetDec~\cite{kvroustek2017retdec}), plus the recent compiler-aware method SAILR~\cite{sailr}.
3) \textbf{Dedicated LLM-based Methods}: Three end-to-end models (LLM4Decompile-End~\cite{tan2024llm4decompile}, Nova~\cite{jiang2023nova}, SccDec~\cite{sccdec}) and one refinement-based model (LLM4Decompile-Ref~\cite{tan2024llm4decompile} which refines Ghidra's output). We use their official HuggingFace weights and default hyperparameters.

\subsubsection{Metrics \& Implementation}
We evaluate decompilation performance using three metrics: 
1) \textbf{Re-Execution Rate (RE)}: Verifies functional correctness by checking if the decompiled code passes all test cases.
2) \textbf{Re-Compilation Rate (RC)}: Assesses syntactic correctness by checking if the code successfully recompiles.
3) \textbf{Test Case Pass Rate (TCP)}: Measures the average proportion of passed test cases across all functions.

\saltm is implemented in $\sim$2000 lines of Python. For disassembly and CFG extraction, we use Angr~\cite{angr} purely for implementation convenience, as our method only requires standard binary analysis capabilities. We use a 6.7B parameter \salte model trained for one epoch (max length 4096, learning rate 5e-6, batch size 8, gradient accumulation 16) in $\sim$8 hours on an Ubuntu 22.04 server with an Intel 48-core CPU and two Nvidia H100 80GB GPUs. DeepSeek-V3 is employed for code optimization.

\subsection{Effectiveness of \saltm (RQ1)}\label{sec:rq1}

In this RQ, we evaluate the effectiveness of \saltm against all baselines on three open-source datasets (Table~\ref{table:rq1}).
Overall, it consistently outperforms all baselines across all metrics.
On the Decompile-Eval dataset, for example, it achieves 58.7\% in re-execution (RE), 96.8\% in re-compilation (RC), and 70.4\% in test case pass (TCP) rates.
Compared to the state-of-the-art end-to-end method, SccDec (49.8\% RE, 92.8\% RC, 59.8\% TCP), \saltm improves RE, RC, and TCP by 8.9\%, 4.0\%, and 10.6\%, respectively.
Notably, the more significant gain in TCP than in RE highlights that \saltm generates higher-quality decompiled code with superior execution correctness and functional equivalence to the original source.
Furthermore, we also demonstrate that even \textbf{without the refinement steps} (Section~\ref{sec:op}) based on the general-purpose LLM, the performance is still superior to any of the current competing baselines (e.g., our \saltm-ALT is 57.2 RE in Table~\ref{table:rq1} on the Decompile-Eval dataset).

% The results reveal distinct performance trends across different baselines.
% Among the general-purpose LLMs, Claude-3.5-sonnet demonstrates impressive capabilities, outperforming most dedicated decompilation methods and achieving results only approximately 10\% below \saltm on the Decompile-Eval dataset.
% This strong performance likely stems from its advanced proficiency in code understanding, particularly in complex programming scenarios.
% In contrast, o1-mini shows limited effectiveness, indicating less well-suited for decompilation tasks. 
% The commercial decompilation tools and SAILR, which are primarily designed to aid human analysts in code comprehension rather then ensure code re-execution, exhibit comparatively weaker performance. 
% Nevertheless, well-established tools such as Hex-Rays and RetDec maintain competitive decompilation quality.
The results also reveal distinct trends among the baselines.
Among general-purpose LLMs, Claude-3.5-sonnet exhibits impressive capabilities, trailing ours by only $\sim$10\% on Decompile-Eval, likely due to its advanced proficiency in resolving coding scenarios.
Conversely, o1-mini shows less suitability for decompilation tasks.
Meanwhile, commercial decompilers and SAILR perform comparatively weaker on these metrics, as they are primarily designed to prioritize human readability over code re-execution, though established tools like Hex-Rays still maintain competitiveness.
Hex-Rays occasionally introduces extra statements for readability, which may explain their lower RE compared to RetDec.
% LLM4Decompile-Ref, which employs an LLM to refine the output of Ghidra, achieved competitive results on Decompile-Eval (with a 52.3\% RE), ranking second only to \saltm and outperforming existing end-to-end methods.
% However, its performance dropped significantly on Exebench, nearly matching the weakest baseline
% %However, its performance on Exebench was substantially weaker, nearly aligning with the lowest-performing baseline.
% This discrepancy aligns with our analysis in Section~\ref{sec:rw} and~\ref{sec:intro}, indicating that refinement-based methods are highly sensitive to decompiler output quality.
% For example, Ghidra’s RE score on Exebench is merely 3.2\%, and LLM4Decompile-Ref’s performance similarly degrades when using Ghidra’s output.
% Ghidra’s pseudo-code may lose critical semantic or structural information during analysis, making it difficult for the LLM to generate correct source code.
% This limitation highlights a key advantage of \saltm’s end-to-end approach: by processing Assembly directly without intermediate decompilation stages, it avoids irreversible information loss inherent in refinement-based methods.
% While other dedicated decompilation methods achieve respectable results, their performance consistently falls short of \saltm. 
% This performance gap originates from their inherent limitation in accurately modeling high-level code logic, which is a challenge that \saltm successfully addresses through its innovative approach of reconstructing \salt from low-level operations.

For refinement-based models, LLM4Decompile-Ref achieves a competitive 52.3\% RE on Decompile-Eval (ranking second only to \saltm), but degrades sharply on Exebench.
As discussed in Sections\ref{sec:rw} and\ref{sec:intro}, this discrepancy stems from its heavy reliance on the underlying decompiler's quality.
For instance, Ghidra generates a mere 3.2\% RE on Exebench, causing irreversible loss of semantic and structural information that strictly bounds the LLM's refinement capability. 
By contrast, \saltm's end-to-end processing of Assembly code completely bypasses this intermediate information loss.
Moreover, although other dedicated LLMs perform decently, \saltm outpaces them by addressing their common inability to model high-level structures, leveraging its novel \salt reconstruction directly from low-level operations.

\begin{tcolorbox}[enhanced, width=\linewidth, boxrule=0.8pt, left=2pt, right=2pt, top=2pt, bottom=2pt, drop fuzzy shadow=black,]
\textbf{Answer to RQ1.} \saltm demonstrates superior performance in decompilation compared to all baseline methods (even removing all refinement steps in Section~\ref{sec:op}). For example, it achieves improvements of 4\% in recompilation rate, 8.9\% in re-execution rate, and 10.6\% in test case pass rate over the SOTA End-to-End decompilation method on the Decompile-Eval dataset. %This reveals that the decompiled results generated by \saltm are of higher quality than those of other baselines.
\end{tcolorbox}

\subsection{Performance under Code Obfuscation (RQ2)}

\begin{table*}[t]
    \centering
    \caption{Performance of Decompilation Methods across Various Obfuscation Techniques}
    \label{table:ob}

    \footnotesize
    \setlength{\tabcolsep}{3.0pt}
    \renewcommand{\arraystretch}{1.08}

    \begin{tabular*}{\textwidth}{
        @{\extracolsep{\fill}}
        l
        ccc
        ccc
        ccc
        ccc
        @{}
    }
        \toprule

        \multirow{2}{*}{\textbf{Model}} &
        \multicolumn{3}{c}{\textbf{Bogus CF}} &
        \multicolumn{3}{c}{\textbf{CF Flattening}} &
        \multicolumn{3}{c}{\textbf{Ins. Substitution}} &
        \multicolumn{3}{c}{\textbf{BB Split}} \\

        \cmidrule(lr){2-4}
        \cmidrule(lr){5-7}
        \cmidrule(lr){8-10}
        \cmidrule(lr){11-13}

        &
        \textbf{RE} & \textbf{RC} & \textbf{TCP} &
        \textbf{RE} & \textbf{RC} & \textbf{TCP} &
        \textbf{RE} & \textbf{RC} & \textbf{TCP} &
        \textbf{RE} & \textbf{RC} & \textbf{TCP} \\

        \midrule

        SccDec-6.7B
        & 18.9 & 39.6 & 24.4
        & 17.1 & 52.0 & 22.0
        & 35.7 & 88.1 & 48.7
        & 37.3 & 88.6 & 50.9 \\

        Nova-6.7B
        & 9.9 & 74.4 & 22.5
        & 6.4 & 64.5 & 15.5
        & 9.8 & 77.4 & 24.9
        & 9.8 & 76.8 & 25.0 \\

        LLM4Decompile-6.7B
        & 17.8 & 39.8 & 22.8
        & 16.2 & 44.7 & 20.3
        & 32.0 & 81.7 & 45.3
        & 33.2 & 83.7 & 45.4 \\

        \textbf{\saltm-6.7B}
        & \textbf{25.5} & \textbf{84.9} & \textbf{36.5}
        & \textbf{20.3} & \textbf{85.1} & \textbf{30.7}
        & \textbf{40.1} & \textbf{95.1} & \textbf{54.3}
        & \textbf{38.3} & \textbf{95.3} & \textbf{53.8} \\

        \bottomrule
    \end{tabular*}
\end{table*}

To assess the robustness of \saltm, we evaluate its performance on binaries obfuscated with various methods.
% We evaluate the performance of \seqsplit{\saltm} when applied to binaries obfuscated using different techniques. 
Code obfuscation is commonly employed in malware to conceal malicious behaviors, complicating analysis and detection.
Following the previous work~\cite{tan2024llm4decompile}, we obfuscate source code from the Decompile-Eval using the widely adopted code obfuscation tool, Obfuscator-LLVM~\cite{junod2015obfuscator}, and employ four standard obfuscation techniques: 
1) \textit{Bogus control flow} (BCF), which inserts fake control flows and pads with garbage instructions to change the program's control flow, 2) \textit{Control flow flattening} (CFF), which reorganizes all basic blocks into a single level structure and encapsulates them within switch statements inside a loop, 3) \textit{Instructions substitution} (IS), which replaces standard instructions with semantically equivalent but more complex instructions, and 4) \textit{Basic block split} (BBS), which splits one basic block into multiple equivalent blocks.

Specifically, to ensure a fair comparison (e.g., input limit, model size and so on), we directly compare \saltm against all LLM-based end-to-end decompilation methods: SccDec, Nova, and LLM4Decompile.
% As shown in Table~\ref{table:ob}, \saltm consistently outperforms all baseline methods across all obfuscation techniques, achieving an average improvement of 16.8\% in recompilation rate, 3.8\% in re-execution rate, and 7.3\% in test case pass rate. 
As shown in Table~\ref{table:ob}, \saltm consistently outperforms all baseline methods across all obfuscation techniques, achieving an average improvement of 16.8\% in RC rate, 3.8\% in RE rate, 
and 7.3\% in TCP rate.
These results highlight the superior capability of \saltm in decompiling obfuscated binaries. 
Among the four obfuscation techniques, CFF and BCF exhibit the most significant impact on \saltm's performance, while IS has the least impact.
This difference likely stems from the fact that CFF and BCF significantly obscure the underlying code logic by extensively altering the control flow, whereas IS mainly modifies individual instructions without affecting the overall program structure. 
% This discrepancy likely arises because CFF and BCF obscure the true logic of the code by drastically altering the control flow, whereas IS primarily modifies individual instructions without disrupting the overall structure.
Specifically, CFF affects performance by transforming loops into switch statements, indicating that incorporating additional logic node types (e.g., $Switch$) into the \salt could further enhance \saltm's effectiveness.
% potential improvements through the incorporation of additional logic node types (e.g., $If$ and $Switch$) into the \salt construction process. 
Notably, \saltm demonstrates its greatest advantage over baselines in BCF, attributed to its ability to recover high-level structures and effectively distinguish fake control flows. In contrast, its performance in BBS obfuscation is comparatively modest, reflecting the challenges of reconstructing \salt after basic block splitting.

\begin{tcolorbox}[enhanced, width=\linewidth, boxrule=0.8pt, 
 left=2pt, right=2pt, top=2pt, bottom=2pt, drop fuzzy shadow=black,]
\textbf{Answer to RQ2.} \saltm achieves superior performance across four code obfuscation techniques, with CFF and BCF exerting the most substantial impact on the performance of all methods. It demonstrates its most significant advantage in BCF but a smaller one in BBS.
\end{tcolorbox}

\subsection{Ablation Study (RQ3)}

We conduct five ablations to evaluate our designs: 1) \salt's structural advantage (Section~\ref{ab_1}), 2) impact of instruction normalization (Section~\ref{sec:ab_in}), 3) other \saltm components (Section~\ref{ab_2}), 4) fixers with different LLMs (Section~\ref{ab_3}), and 5) \salt's scalability across model sizes (Section~\ref{ab_4}).

\subsubsection{Structural Advantage of \salt}\label{ab_1}
We evaluate three variants (+Linear, +CFG~\cite{liupeipei}, +\salt) using DeepSeekCoder and LLM4Decompile.
All share the same training data and preprocessing pipeline to strictly control fair comparisons.
Crucially, because all variants are identically equipped with IN, any performance gap is derived from different binary representation.
Table~\ref{table:ablation_salt} shows \textbf{+\salt} (e.g., 57.2\% RE on LLM4Decompile-6.7B) significantly outperforms \textbf{+Linear} (46.7\% RE) and \textbf{+CFG} (51.8\% RE) across all bases. This compellingly confirms that \salt's loop-oriented structure provides an obvious advantage over simple instruction-level operations, effectively guiding LLMs via shared logic patterns (Section~\ref{sec:mt}). Conversely, CFG models show inconsistent gains (sometimes underperforming +Linear), due to the LLMs' limited capacity to infer high-level loops from raw control flows and limited training data.

\begin{table*}[t]
    \centering
    \caption{Effectiveness of \salt Compared with Different Binary Representations}
    \label{table:ablation_salt}

    \begin{threeparttable}
    \footnotesize
    \setlength{\tabcolsep}{3pt}
    \renewcommand{\arraystretch}{1.08}

    \begin{tabular*}{\textwidth}{
        @{\extracolsep{\fill}}
        l
        ccccc
        ccccc
        c
        @{}
    }
        \toprule

        \multirow{2}{*}{\textbf{Model}} &
        \multicolumn{5}{c}{\textbf{Re-execution Rate}} &
        \multicolumn{5}{c}{\textbf{Re-compilation Rate}} &
        \multirow{2}{*}{\textbf{TCP Rate}} \\

        \cmidrule(lr){2-6}
        \cmidrule(lr){7-11}

        &
        \textbf{O0} & \textbf{O1} & \textbf{O2} & \textbf{O3} & \textbf{Avg.} &
        \textbf{O0} & \textbf{O1} & \textbf{O2} & \textbf{O3} & \textbf{Avg.} &
        \\

        \midrule

        \textbf{DeepSeekCoder-6.7B}
        & 0.0 & 0.0 & 0.0 & 0.0 & 0.0
        & 1.2 & 0.0 & 0.0 & 1.8 & 0.8
        & 0.0 \\

        \hspace{0.6cm}+ Linear$^\ast$
        & 51.2 & 23.8 & 28.0 & 26.8 & 32.5
        & 89.8 & 79.9 & 78.5 & 80.2 & 82.1
        & 43.2 \\

        \hspace{0.6cm}+ CFG$^\ast$
        & 52.4 & 24.2 & 28.1 & 26.6 & 32.8
        & 90.1 & 79.1 & 78.7 & 80.5 & 82.1
        & 44.7 \\

        \hspace{0.6cm}+ ALT$^\ast$
        & \textbf{57.9} & \textbf{30.5} & \textbf{31.1} & \textbf{31.1} & \textbf{37.7}
        & \textbf{90.2} & \textbf{79.9} & \textbf{79.3} & \textbf{81.1} & \textbf{82.6}
        & \textbf{49.4} \\

        \midrule

        \textbf{LLM4Decompile-6.7B}
        & 67.7 & 41.5 & 40.9 & 36.6 & 46.7
        & 92.7 & 85.4 & 87.2 & 86.6 & 88.0
        & 57.9 \\

        \hspace{0.6cm}+ Linear$^\ast$
        & 66.5 & 48.2 & 45.7 & 41.5 & 50.5
        & 92.1 & \textbf{88.6} & 90.2 & 89.6 & 90.1
        & 60.4 \\

        \hspace{0.6cm}+ CFG$^\ast$
        & 70.1 & 49.3 & 46.2 & 41.6 & 51.8
        & 93.3 & 88.3 & 90.4 & 89.9 & 90.5
        & 62.1 \\

        \hspace{0.6cm}+ ALT$^\ast$
        & \textbf{82.3} & \textbf{51.8} & \textbf{48.8} & \textbf{45.7} & \textbf{57.2}
        & \textbf{95.7} & 88.4 & \textbf{90.9} & \textbf{90.2} & \textbf{91.3}
        & \textbf{67.1} \\

        \hspace{0.6cm}+ \salt (w/o IN)
        & 68.9 & 49.4 & 47.0 & 43.9 & 52.3
        & 95.7 & 88.3 & 90.8 & 90.2 & 91.2
        & 64.6 \\

        \midrule

        \textbf{LLM4Decompile-1.3B}
        & 47.6 & 25.6 & 23.8 & 20.7 & 29.4
        & 87.5 & 82.9 & 72.3 & 72.3 & 78.8
        & 42.5 \\

        \hspace{0.6cm}+ ALT$^\ast$
        & \textbf{62.8} & \textbf{33.5} & \textbf{28.1} & \textbf{28.0} & \textbf{38.1}
        & \textbf{89.6} & \textbf{84.1} & \textbf{76.2} & \textbf{75.6} & \textbf{81.4}
        & \textbf{50.1} \\

        \bottomrule
    \end{tabular*}

    \begin{tablenotes}
        \footnotesize
        \item $\ast$ All fine-tuned variants include IN for fair comparison.
        +CFG follows the method in CodeInverter~\cite{liupeipei}.
    \end{tablenotes}

    \end{threeparttable}
\end{table*}

\begin{table*}[t]
    \centering
    \caption{Ablation Results on VR, CEF, and BEF of \saltm}
    \label{table:ablation_com}

    \begin{threeparttable}
    \footnotesize
    \setlength{\tabcolsep}{3pt}
    \renewcommand{\arraystretch}{1.08}

    \begin{tabular*}{\textwidth}{
        @{\extracolsep{\fill}}
        l
        ccccc
        ccccc
        c
        @{}
    }
        \toprule

        \multirow{2}{*}{\textbf{Model}} &
        \multicolumn{5}{c}{\textbf{Re-execution Rate}} &
        \multicolumn{5}{c}{\textbf{Re-compilation Rate}} &
        \multirow{2}{*}{\textbf{TCP Rate}} \\

        \cmidrule(lr){2-6}
        \cmidrule(lr){7-11}

        &
        \textbf{O0} & \textbf{O1} & \textbf{O2} & \textbf{O3} & \textbf{Avg.} &
        \textbf{O0} & \textbf{O1} & \textbf{O2} & \textbf{O3} & \textbf{Avg.} &
        \\

        \midrule

        \textbf{\saltm}
        & \textbf{80.5}
        & \textbf{52.4}
        & 53.7
        & \textbf{48.2}
        & \textbf{58.7}
        & \textbf{98.2}
        & \textbf{97.6}
        & \textbf{97.0}
        & 94.5
        & 96.8
        & \textbf{70.4} \\

        \hspace{0.6cm}-- w/o VR
        & 76.5
        & 49.9
        & 50.4
        & 44.8
        & 55.4
        & 96.3
        & 96.1
        & 94.7
        & 91.2
        & 94.6
        & 65.6 \\

        \hspace{0.6cm}-- w/o BEF
        & 82.9
        & 52.4
        & 50.0
        & 47.6
        & 58.2
        & 98.2
        & 97.6
        & 97.0
        & \textbf{95.1}
        & \textbf{97.0}
        & 69.2 \\

        \hspace{0.6cm}-- w/o CEF
        & 78.7
        & 51.8
        & \textbf{56.7}
        & 47.6
        & 58.7
        & 95.7
        & 92.7
        & 95.1
        & 93.3
        & 94.2
        & 70.1 \\

        \bottomrule
    \end{tabular*}

    \begin{tablenotes}
        \footnotesize
        \item \textbf{VR}: variable renaming;
        \textbf{BEF}: boundary error fixer;
        \textbf{CEF}: compilation error fixer.
    \end{tablenotes}

    \end{threeparttable}
    \vspace{-0.3cm}
\end{table*}

\subsubsection{Impact of Instruction Normalization (IN)}\label{sec:ab_in}
We test the variant \textbf{+ \salt (w/o IN)} to explicitly evaluate IN's contribution compared to \textbf{+ \salt} (also means the performance of \salte model). Removing IN degrades the model's RE from 57.2\% to 52.3\%, validating its role in supplying LLMs with crucial data references like embedded constants (Section~\ref{sec:mt}). Notably, despite lacking normalization, the unoptimized \textbf{+ \salt (w/o IN)} still decisively beats the IN-equipped \textbf{+Linear} (50.5\%) and \textbf{+CFG} (51.8\%), reiterating the sheer dominance of \salt's structural superiority.

\subsubsection{Impact of VR, CEF and BEF}\label{ab_2}
As shown in Table~\ref{table:ablation_com}, excluding variable renaming (VR) during the training stage drops the RE from 58.7\% to 55.4\%, validating our strategy of prioritizing execution semantics over symbol optimizations~\cite{hu2024degpt}. Similarly, removing the compilation error fixer (CEF) predictably degrades the RC. Interestingly, under \texttt{O2} optimization, CEF slightly reduces RE; our analysis reveals that in these rare compiler-induced cases, the boundary error fixer (BEF) handles repairs more effectively. Overall, \salt remains independently powerful and complementary to these modular enhancements.

% \begin{table}[t]
% \centering
% %\vspace{-0.5em}
% \renewcommand{\arraystretch}{1.0}
% \setlength {\tabcolsep} {3pt}
% \caption{Ablation Results on Fixers with Different LLMs}
% %\vspace{-0.9em}
% \begin{threeparttable}
% \resizebox{0.65\textwidth}{!}{%
% \begin{tabular}{lccc}
% \toprule
% \multicolumn{1}{c}{\textbf{Model}} & \textbf{RE (AVG.)} & \textbf{RC (AVG.)} & \textbf{TCP rate} \\
% \midrule
% \textbf{\saltm} & 58.7 & \textbf{96.8} & 70.4 \\
% \hspace{0.2cm} - w/ o1\_mini & 59.9 & 96.0 & 70.5 \\ 
% \hspace{0.2cm} - w/ GPT\_4o & 58.1 & 96.3 & 69.5 \\
% \hspace{0.2cm} - w/ Claude\_3.5 & \textbf{60.8} & \textbf{96.8} & \textbf{72.2}  \\
% \bottomrule
% \end{tabular}
% }
% %\vspace{-1em}
% \end{threeparttable}
% \label{table:albation_fixer}
% \end{table}

\begin{table}[t]
    \centering
    \caption{Ablation Results on Fixers with Different LLMs}
    \label{table:albation_fixer}

    \footnotesize
    \setlength{\tabcolsep}{3pt}
    \renewcommand{\arraystretch}{1.08}

    \begin{tabular*}{\columnwidth}{
        @{\extracolsep{\fill}}
        lccc
        @{}
    }
        \toprule
        \textbf{Model} &
        \textbf{RE (Avg.)} &
        \textbf{RC (Avg.)} &
        \textbf{TCP Rate} \\
        \midrule

        \textbf{\saltm}
        & 58.7
        & \textbf{96.8}
        & 70.4 \\

        \hspace{0.15cm}w/ o1-mini
        & 59.9
        & 96.0
        & 70.5 \\

        \hspace{0.15cm}w/ GPT-4o
        & 58.1
        & 96.3
        & 69.5 \\

        \hspace{0.15cm}w/ Claude-3.5
        & \textbf{60.8}
        & \textbf{96.8}
        & \textbf{72.2} \\

        \bottomrule
    \end{tabular*}
    \vspace{-0.6cm}
\end{table}

\subsubsection{Impact of different Fixer}\label{ab_3}

We evaluate the impact of different fixers implemented using various LLMs. Specifically, we select three representative LLMs and present the results in Table~\ref{table:albation_fixer}. 
It can be observed that Claude fixer achieves the highest overall performance. 
However, due to API usage cost consideration, we opt to use a more affordable model as the default fixer in \saltm.
% for the current fixer. 
Although o1-mini shows unsatisfactory performance in decompilation tasks, it ranks second in decompilation code fixing. This observation indirectly highlights the importance of prioritizing the accurate recovery of execution semantics during decompilation.
% of completing the correct recovery of execution semantics.

\subsubsection{Performance across different Model Scale}\label{ab_4}
We further assess whether \salt maintains its advantages on smaller-scale models.
As indicated in Table~\ref{table:ablation_salt}, \salt brings a significant improvement to both 1.3B and 6.7B models. Notably, the re-execution rate increases by 8.7\% for the 1.3B model (from 29.4\% to 38.1\%) and by 10.5\% for the 6.7B model (from 46.7\% to 57.2\%). 
These results confirm that \salt remains highly effective across varying model scales.
%By refactoring Assembly code, \salt guides LLMs in understanding intricate execution logic.

\begin{tcolorbox}[enhanced, width=\linewidth, boxrule=0.8pt, 
 left=2pt, right=2pt, top=2pt, bottom=2pt, drop fuzzy shadow=black,]
\textbf{Answer to RQ3.} Each component in \saltm contributes significantly to its overall performance, with \salt itself proving instrumental in enhancing decompilation capabilities across diverse model types and scales.
\end{tcolorbox}

\subsection{Real-world Applications (RQ4)}

\begin{figure}[h]
    \centering
    %\vspace{-2em}
    \includegraphics[width=0.8\linewidth]{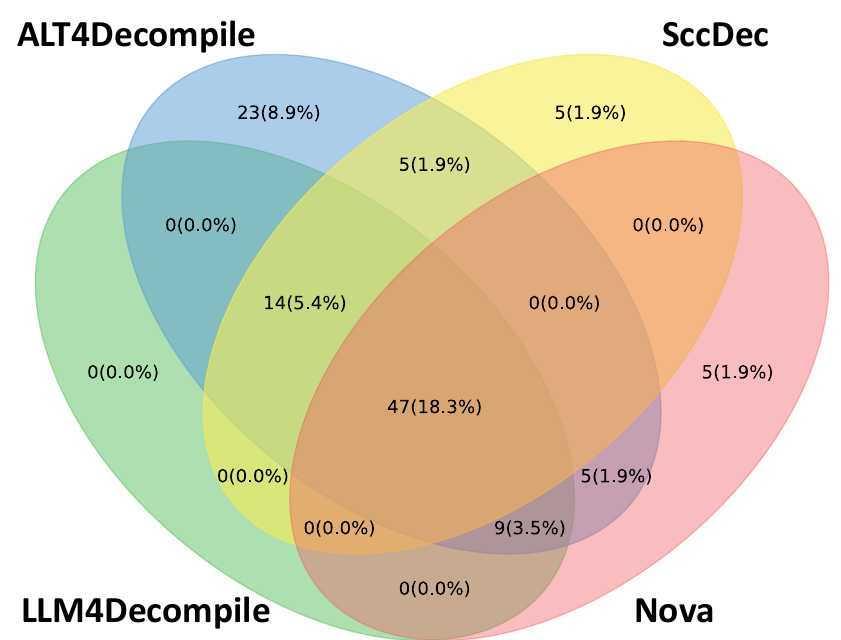}
    \caption{Real-world Applications.}
   % \Description{}
    \label{fig:ra}
    %\vspace{-1.0em}            
\end{figure}

We compare \saltm against LLM-based end-to-end baselines on real-world software to ensure a fair comparison (e.g., input limit, model size and so on).
Since real-world binaries lack test I/O pairs, we use an Assembly-search dataset~\cite{gao2024virtual} comprising 257 binary files (e.g., \textit{WeChatWin.dll}) paired with natural language descriptions. After decompiling all functions, we use DeepSeek-V3 to summarize the outputs. Human evaluators then assess accuracy by comparing these summaries and the original Assembly against ground-truth descriptions.
As Figure~\ref{fig:ra} shows, \saltm correctly decompiles the most functions (102, 39.7\%), while Nova performs worst (66, 25.7\%). Furthermore, 47 cases are solved by all methods, but \saltm achieves the most unique successes (23 cases).

\smallskip \noindent \textbf{Error Analysis}. We identify three primary failure causes.
First, all models frequently generate meaningless assignments (highest in SccDec at 48.9\%, lowest in Nova at 34.0\%). We attribute this to imbalanced training data in the base LLM4Decompile, which overemphasizes assignment patterns from Exebench (Section~\ref{sec:rn}). While our data filtering and Nova's diverse training mitigated this, the bias persists. 
Second, the typical 4096-token context window of LLM-based methods truncates the decompilation of long real-world functions, causing incomplete outputs across all models.
Finally, obfuscation triggers analysis tool failures (e.g., disassembly errors).

\begin{tcolorbox}[enhanced, width=\linewidth, boxrule=0.8pt, 
 left=2pt, right=2pt, top=2pt, bottom=2pt, drop fuzzy shadow=black,]
\textbf{Answer to RQ4.} \saltm achieves the highest accuracy (39.7\%, a 12.1\% improvement) on real-world software. However, inherent LLM context limits and analysis tool dependencies remain key bottlenecks limiting further gains.
\end{tcolorbox}

% \begin{figure}
%     \centering
%     %\vspace{-2em}
%     \includegraphics[width=0.95\linewidth]{figs/2-figure.pdf}
%    \vspace{-1.0em}   
%     \caption{Results of Real-world Applications and User Study.}
%    % \Description{}
%     \label{fig:us}
%     \vspace{-1.5em}            
% \end{figure}

% \begin{figure}[t]
%     \centering
%     \begin{minipage}[t][][t]{0.44\textwidth}
%         \centering
%         \includegraphics[trim=0 -50 0 0, width=\textwidth]{figs/venn.pdf}
%         %\vspace{-5em}
%         \caption{Real-world Applications.}
%         \label{fig:ra}
%     \end{minipage}
%     \hfill
%     \begin{minipage}[t][][t]{0.5\textwidth}
%         \centering
%         \includegraphics[width=\textwidth]{figs/figure.pdf}
%        % \vspace{-2em}
%         \caption{User Study.}
%         \label{fig:us}
%     \end{minipage}
%     %\vspace{-2em}
% \end{figure}

\subsection{Cost Study (RQ5)}
\begin{figure}[h]
    \centering
    %\vspace{-2em}
    \includegraphics[width=0.8\linewidth]{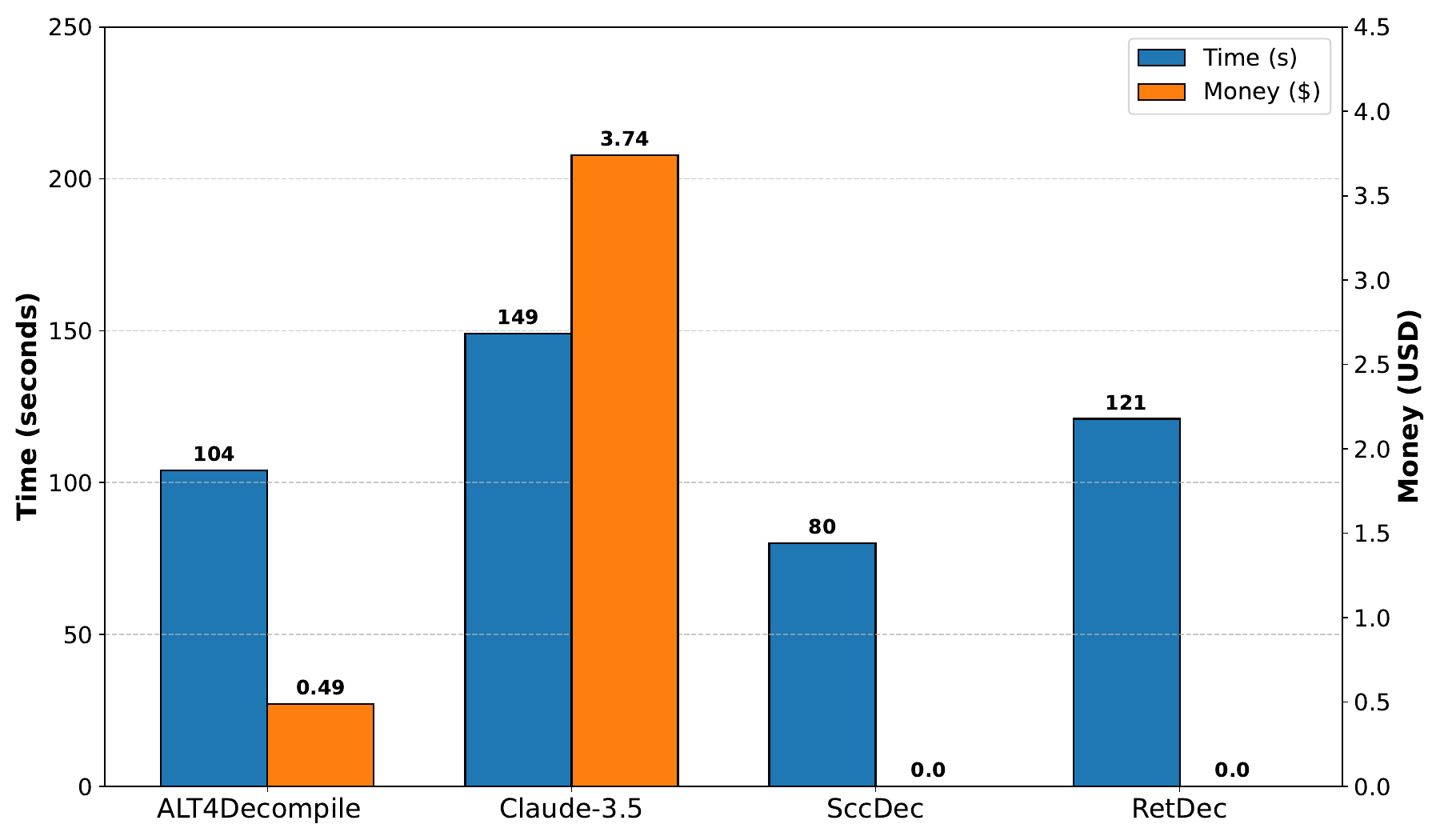}
    \caption{Time and money cost of different methods.}
   % \Description{}
    \label{fig:ct}
    %\vspace{-1.0em}            
\end{figure}

In this RQ, we evaluate how much time and money \saltm and other methods cost on the Decompile-Eval dataset. 
We select the best-performing baselines (Claude, RetDec, SccDec) from the three baseline categories for comparison.
The result is shown in Figure~\ref{fig:ct}.
Specifically, the decompilation time of \saltm is only 7 seconds longer than that of SccDec, with the additional time attributed to CFG extraction and \salt construction.
Moreover, \saltm incurs an additional time cost of 17 seconds for the fixer and symbolic information recovery phases. Among all methods,
Claude is the slowest in getting feedback.
In terms of monetary cost, \saltm incurs only \$0.49, which is significantly lower than Claude's \$3.74.
Notably, all of \saltm's cost is attributed to symbolic information recovery and a small portion of error correction. 
As shown in Table~\ref{table:ablation_salt}, even without fixers and symbolic information recovery, 
\salt achieves a RE rate of 0.572 (only 0.015 lower than 0.587) on the LLM4Decompile model, which is still much higher than the other baseline models.
Therefore, the additional \$0.49 and 17s can be considered optional when resources are constrained, allowing users to make a practical trade-off between performance and cost.

\begin{tcolorbox}[enhanced, width=\linewidth, boxrule=0.8pt, 
 left=2pt, right=2pt, top=2pt, bottom=2pt, drop fuzzy shadow=black,]
\textbf{Answer to RQ5.} \saltm does not bring about excessive additional costs. The necessary additional time costs are CFG extraction and \salt construction (7s on the Decompile-Eval dataset).
\end{tcolorbox}

\subsection{User Study}

\begin{figure}[h]
    \centering
    %\vspace{-2em}
    \includegraphics[width=1.0\linewidth]{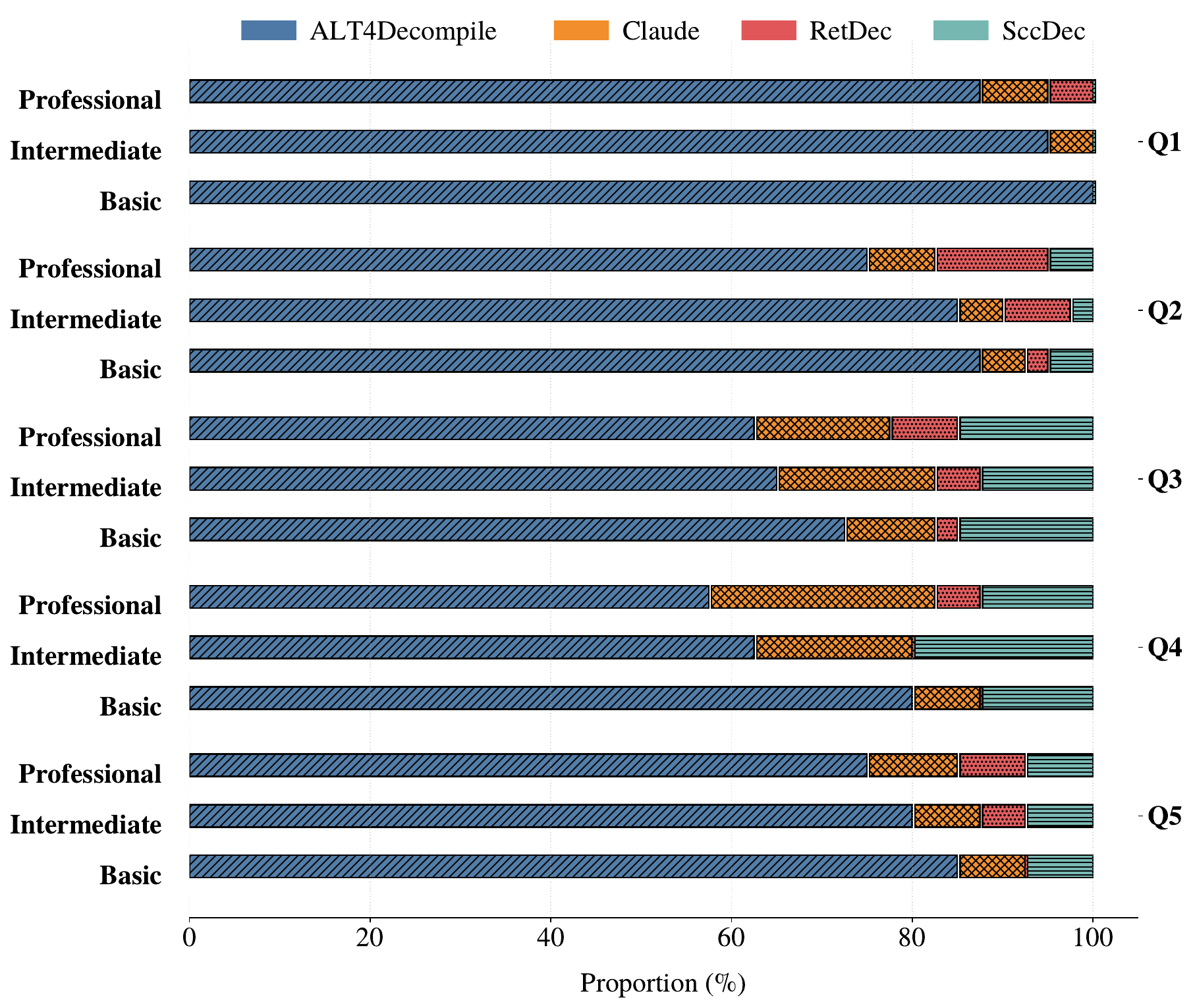}
    \caption{User Study.}
   % \Description{}
    \label{fig:us}
    %\vspace{-1.0em}            
\end{figure}

% \begin{figure*}[t]
%     \centering
%     \includegraphics[width=0.8\textwidth]{figs/figure.pdf}
%     \caption{User Study.}
%     \label{fig:us}
% \end{figure*}

\saltm aims to produce decompiled code that (1) enables accurate program comprehension by reverse engineers, (2) preserves functional correctness for direct execution, and (3) minimizes the effort required for practical reuse.
Drawing from previous work~\cite{hu2024degpt}, we conduct a user study comprising five questions to assess \saltm's effectiveness in achieving this goal.
Specifically, we recruit 12 participants, grouped by experience level: Professional (with over four years of reverse engineering experience), Intermediate (with 1-2 years of reverse engineering experience), and Basic (with an understanding of basic reverse engineering concepts and familiarity with code debugging processes).
We compare \saltm against the top-performing baselines from three categories: Claude, RetDec, and SccDec.
% We select the best-performing baselines (Claude, RetDec, SccDec) from the three baseline categories for comparison.
Each participant is presented with 10 randomly selected functions, along with the corresponding source code, test cases, and four decompiled outputs, anonymized to prevent bias. 
For each function, participants answer five evaluation questions. In total, we collect 120 responses per question. The questions are as follows:

\begin{itemize}[leftmargin=10pt]
    \item \textbf{Q1:} Which decompiled code contains the most meaningful and helpful comments? 
    % Which code's comments do you think are more meaningful and helpful in all the decompiled options?
    \item \textbf{Q2:} Which decompiled code has the most meaningful and helpful variable names?
    % Which code's variables do you think are more meaningful and helpful in all the decompiled options?
    \item \textbf{Q3:} Which decompiled code best preserves the original function's semantics or behavior?
    % Which code best restores the original function's semantics (or behavior) in all the decompiled options?
    \item \textbf{Q4:} Which decompiled code requires the fewest debugging steps to pass all provided test cases?
    % Which code needs the least debugging steps to pass all the test cases in the test code in all the decompiled options?
    \item \textbf{Q5:} Overall, which decompiled code helps you best understand the function’s logic, compared to the original version?
    % Overall, which code helps you better understand the function compared to the original function?
\end{itemize}

% \begin{figure}
%     \centering
%     %\vspace{-2em}
%     \includegraphics[width=0.75\linewidth]{figs/figure.pdf}
%    \vspace{-0.6em}   
%     \caption{Results of User Study.}
%    % \Description{}
%     \label{fig:us}
%     %\vspace{-0.5em}            
% \end{figure}

As Figure~\ref{fig:us} shows, \saltm consistently receives more positive feedback across all five questions.
Notably, the Basic group found some original source codes difficult to comprehend but achieved better understanding using \saltm's outputs. Conversely, consistent with DeGPT's findings, the Professional group provided comparatively lower ratings; their extensive expertise naturally reduces reliance on LLM-generated readability enhancements, allowing them to effectively interpret raw commercial decompiler outputs. Furthermore, \saltm's highest ratings in Q4 confirm its superior semantic preservation, directly corroborating our quantitative execution results in Section~\ref{sec:rq1}.

\begin{tcolorbox}[enhanced, width=\linewidth, boxrule=0.8pt, 
 left=2pt, right=2pt, top=2pt, bottom=2pt, drop fuzzy shadow=black,]
\textbf{Answer to User Study.} Overall, \saltm receives the most votes from the 12 participants, with the Basic group benefiting the most from our method. In addition, most participants agree that even if \saltm produced incorrect results, they could be corrected with minimal debugging effort.
\end{tcolorbox}

\section{Related Work}\label{sec:rw}
% Traditional decompilers (e.g., Hex-Rays~\cite{IDApro}) rely on program analysis.
% With advancements in deep learning, many studies employ neural machine translation (Transformers~\cite{vaswani2017attention}) to transform binary into high-level code.
% Based on their research focus, we classify these methods into three categories.
Traditional decompilers~\cite{IDApro} rely on program analysis, while recent advances leverage neural translation~\cite{vaswani2017attention} to map binaries to high-level code. We categorize existing techniques into three groups:

\smallskip \noindent \textbf{Rule-based methods.} 
% These methods transform low-level code into readable C-like pseudocode through static analysis.
% Hex-Rays~\cite{IDApro}, Ghidra~\cite{ghidra} decompilers apply heuristics to recover types and simplify expressions.
% Phoenix~\cite{Phoenix}, DREAM~\cite{Dream}, and SAILR~\cite{sailr} introduce different structuring algorithms to eliminate gotos.
% However, they often rely heavily on costly expert knowledge and manual intervention.
% We attempt to address this by following a simple idea: high- and low-level languages share similar logic features that can be systematically designed and extracted with minimal human effort, thereby enabling LLMs to perform more accurate decompilation.
Tools like Hex-Rays~\cite{IDApro} and Ghidra~\cite{ghidra} use static analysis and heuristics for type recovery and expression simplification. Meanwhile, Phoenix~\cite{Phoenix}, DREAM~\cite{Dream}, and SAILR~\cite{sailr} employ structuring algorithms to eliminate \texttt{goto}s. Despite their utility, they rely heavily on costly expert knowledge and manual intervention. To address this, we systematically extract shared logic features between high- and low-level languages with minimal human effort, guiding LLMs toward more accurate decompilation.

\smallskip \noindent \textbf{Refine-based methods.} 
% These methods utilize LLMs to optimize decompiler outputs.
% DIRE~\cite{dire} employs pseudocode structural information to assist in variable renaming, while DIRTY~\cite{dirty} integrates data layout analysis to predict variable types.
% DecGPT~\cite{decgpt} refines pseudocode by providing LLMs with memory error feedback.
% DeGPT~\cite{hu2024degpt} introduces a three-role mechanism to enhance pseudocode readability.
% ReSym~\cite{xie2024resym} fine-tunes two LLMs to separately restore variables and custom structures.
% DecLLM~\cite{wong2025decllm} combines dynamic runtime feedback with LLMs to improve recompilation rate. 
% However, they rely heavily on the quality of the decompiler outputs, neglecting the overhead of decompiler extensions and the unverified efficacy of the models on raw binary files.
% Consequently, this paper focuses on Assembly code as the main research subject.
These approaches use LLMs to optimize decompiler outputs.
Examples include improving variable renaming and type prediction (DIRE~\cite{dire}, DIRTY~\cite{dirty}, ReSym~\cite{xie2024resym}), or enhancing readability and recompilability via error feedback mechanisms (DecGPT~\cite{decgpt}, DeGPT~\cite{hu2024degpt}, DecLLM~\cite{wong2025decllm}).
However, they rely heavily on the quality of the decompiler outputs, neglecting the overhead of decompiler extensions and the unverified efficacy of the models on raw binary files.
Consequently, this paper focuses on Assembly code as the main research subject.

\smallskip \noindent \textbf{End-to-end methods.}
%These methods treat Assembly code as a linear sequence and utilize a translation model to decompile it.
These methods directly utilize a translation model to decompile Assembly code.
TRAFIX~\cite{katz2019towards} and BTC~\cite{hosseini2022beyond} employ RNNs and Transformers to improve accuracy.
Slade~\cite{armengol2024slade} utilizes PsycheC~\cite{melo2017inference} to infer and supplement missing types.
With the rise of LLMs, Nova~\cite{jiang2023nova} and LLM4Decompile~\cite{tan2024llm4decompile} perform end-to-end training on large-scale source-Assembly pairs.
SccDec~\cite{sccdec} introduces a fine-grained alignment enhancement method (FAE) to improve LLM fine-tuning, thus improving the decompilation capabilities of LLMs.
CodeInverter~\cite{liupeipei} introduces control flow graphs in language model training with millions of functions to improve decompilation.
However, these methods still simply treat Assembly code as a new high-level language to process and attempt to solely rely on the strong learning ability of LLMs to complete decompilation, thereby resulting in no substantial performance gains with limited training (e.g., +CFG in experiments (Section~\ref{ab_1})).
% In our experiments (Section~\ref{ab_1}), the inclusion of CFGs during training yielded no substantial performance gains, indicating that LLMs have difficulty comprehending CFG representations with limited data.
% Therefore, LLMs still struggle to effectively handle instruction streams with arbitrary control flow jumps (in Section~\ref{sec:intro}).

In this paper, inspired by prior research~\cite {m1,jiang2024binaryai,m3}, we observe that some logic patterns shared between high- (C) and low-level (Assembly) languages can effectively guide the decompilation LLM.
Therefore, we propose a novel method for abstracting the high-level loop tree (\salt) from Assembly code, which is different from the existing methods that treat Assembly as a simple language sequence. 
Our \salt significantly improves the performance of LLM-based decompilation.
Notably, unlike Wadec~\cite{she2024wadec}, which performs step-by-step decompilation by directly slicing the inherent loops in Wasm, we refactor C Assembly (which lacks logic structures such as loops) into \salt based on specific jump patterns, and inject it into the LLM all at once for decompilation.

\section{Discussion}

\noindent \textbf{Applicable scenarios.}
Our \salt introduces a high-level loop framework inferred from shared logic patterns between high- (C) and low-level (Assembly) languages.
However, this advantage diminishes when applied to functions comprising only simple assignments, as such functions generally pose less challenge for language models.
In contrast, our method remains more effective in handling complex functions.
The benefit may also decrease if certain analysis tools (e.g., disassemblers) or algorithms for identifying jumping units fail (e.g., wrong entry/exit nodes in connected subgraphs).

\noindent \textbf{Future work.} We explore the effectiveness of selecting the loop structure as a ALT node, which brings significant decompilation LLM performance improvement.
It dues to its significant impact on decompilation failures of existing LLM-based methods (shown in Section~\ref{sec:mt}).
Future work will incorporate other high-level structures such as conditional statements ($if$) to enhance \salt's robustness against diverse code obfuscations and optimizations.

\smallskip \noindent \textbf{Selected evaluation metrics.} 
Previous methods often relied on similarity metrics (e.g., edit distance) to evaluate decompilation quality.
However, these metrics have notable limitations.
First, a low similarity score does not necessarily indicate incorrectness, as the same functionality can be implemented using different logic structures (e.g., a $for$ loop can be replaced by a $while$ loop).
Second, a high similarity score does not guarantee correctness, since even a minor change (e.g., flipping the comparison operator $>$ to $<$ in an $if$ node) can alter the entire execution path of the program.
As discussed in Section~\ref{sec:intro}, this paper focuses on restoring the execution logic correctness of source code.
Therefore, we evaluate the functional correctness of the decompiled results through test case execution (RE and TCP), ensuring that the decompiled code is not only logically correct but also directly reusable.
These metrics provide a more rigorous and meaningful assessment of decompilation quality than traditional superficial similarity-based metrics.

%\vspace{-1.5em}
\section{Conclusion}

In this paper, we propose \saltm, an LLM-based binary decompilation technique.
Unlike previous methods that directly process Assembly as another high-level language, we infer a \salt from binaries, thereby mitigating the syntactic gap between high- and low-level code. 
Evaluation on three datasets demonstrates the effectiveness of \saltm, i.e., showing a 4\% improvement in re-compilation rate, an 8.9\% improvement in re-execution rate, and a 10.6\% improvement in test case pass rate compared to SOTA methods on the Decompile-Eval.
We also conduct a real-world software analysis and user study to further evaluate \saltm.

% ============================================================
% References
% ============================================================

\bibliographystyle{IEEEtranS}
\bibliography{IEEEabrv,ref}

\end{document}